\documentclass[12pt]{article}
\usepackage{epsfig,epsf}
\usepackage{amsmath}
\usepackage{amsfonts}
\usepackage{latexsym}
\usepackage{color}
\newcommand{\be}{\begin{equation}}
\newcommand{\ee}{\end{equation}}
\newcommand{\bea}{\begin{eqnarray}}
\newcommand{\beas}{\begin{eqnarray*}}
\newcommand{\eea}{\end{eqnarray}}
\newcommand{\eeas}{\end{eqnarray*}} 
\newcommand{\ba}{\begin{array}}
\newcommand{\ea}{\end{array}}

\newcommand{\fr}[2]{\frac{#1}{#2}}

\newcommand{\mk}{m_{KK}}   %This can be changed depending on what notation we 
\def\ga{\mathrel{\raise.3ex\hbox{$>$\kern-.75em\lower1ex\hbox{$\sim$}}}}
\def\la{\mathrel{\raise.3ex\hbox{$<$\kern-.75em\lower1ex\hbox{$\sim$}}}}

\def\ls{\mathrel{\lower4pt\vbox{\lineskip=0pt\baselineskip=0pt
           \hbox{$<$}\hbox{$\sim$}}}}
\def\gs{\mathrel{\lower4pt\vbox{\lineskip=0pt\baselineskip=0pt
           \hbox{$>$}\hbox{$\sim$}}}}

\begin{document}

 \begin{flushright}
hep-ph/0305010
\\
TRI-PP-03-07
\\
UVIC-TH/05-03
\end{flushright}

\begin{center}

{\bf 
Cosmological bounds on large extra dimensions from non-thermal production of 
Kaluza-Klein modes
}\\

\vspace{1cm}

Rouzbeh Allahverdi~$^{1}$, Chris Bird~$^{2}$, 
Stefan Groot Nibbelink~$^{2}$ \\ and
Maxim Pospelov~$^{2,3}$

\vspace{1cm}

$^{1}$ {\it Theory Group, TRIUMF, 4004 Wesbrook Mall, \\
Vancouver, B.C., V6T 2A3, Canada.} \\[1ex]

$^{2}$ {\it Department of Physics and Astronomy, University of
Victoria, \\ 
 Victoria, B.C., V8P 1A1, Canada.} \\[1ex]

$^{3}$ {\it Centre for Theoretical Physics, University of Sussex, \\
Brighton BN1 9QJ, U.K.} \\
\vspace{1cm}

\end{center}
\begin{abstract}
\noindent
The existing cosmological constraints on theories with large extra
dimensions rely on the thermal production of the Kaluza-Klein modes  
of gravitons and radions in the early Universe. Successful inflation
and reheating, as well as baryogenesis, typically requires the
existence of a TeV-scale field in the bulk, most notably the inflaton.
The non-thermal production of  KK modes with masses of order 
$100~\rm GeV$  accompanying the inflaton decay sets the lower 
bounds on the fundamental scale $M_*$. For a $1$ TeV inflaton, the late decay 
of these modes distort the successful predictions of Big Bang Nucleosynthesis 
unless $M_*> 35,~13 ,~ 7, ~5$ and $3$ TeV for  $2,~3,~4, ~5$ and $6$ extra
dimensions, respectively. This improves the existing bounds from cosmology on 
$M_*$ for 4, 5 and 6 extra dimensions. Even more stringent bounds are derived 
for a heavier inflaton.

\end{abstract}

\newpage

\section{Introduction}
\label{sc:intro}

The original motivation for theories with large extra spatial 
dimensions was the explanation of the hierarchy between the 
electroweak scale and the Planck mass. In this set up our
four-dimensional  world is assumed to be a flat hyper surface, called
the Standard Model (SM)  brane, which is embedded in a higher
dimensional space-time, known as the bulk. The weakness of  gravity in
$3+1$ dimensions is ascribed to the dilution of the gravitational
interaction by the large volume of the extra dimensions transparent
for gravity, but inaccessible to SM matter and gauge fields. The
hierarchy problem can be solved by assuming that the fundamental scale
$M_*$ in the $4+d$ dimensional bulk is of the order of a
TeV~\cite{nima0,early,review}. However, this requires the extra
dimensions to be large. More quantitative, let $R$ set the 
{\em common} scale of the $d$ extra compact dimensions such  that
their volume is $(2 \pi R)^d$. Then the effective four-dimensional Planck mass
becomes  
\begin{equation} \label{volume}
M_{\rm p}^2 = M_{*}^{2+d} R^d\,.
\end{equation}
This dilution of gravity results in modifications of Newton's law
at distances smaller than $R$. Take $M_{*}=1$ TeV for example, then
for one extra dimension, $R$ will be of order of the size of solar system, 
which is clearly ruled out. However, already for $d=2$ one finds a
value of $R\simeq 0.2$ mm, which is in reach of current experimental 
searches for deviations from Newtonian gravity~\cite{exp2}.

From an effective four dimensional perspective, the distinctive
feature of large extra dimension models is the appearance of a tower
of massive graviton states, known as Kaluza-Klein (KK) graviton modes. 
The perturbations of components of the $4+d$ dimensional metric tensor
behave as a rank $2$-tensor (the graviton),  vectors (the
gravi-photons), and scalars (gravi-scalars) respectively. The diagonal
part of the gravi-scalar matrix contains the $d$ so-called radions, 
which couple to the trace of the four dimensional energy momentum
tensor. (For details see Ref.~\cite{decom}.) 
The energy of these modes is given by 
\(
E^2 = k^2 + \sum_{i} n^{2}_{i} {R}^{-2}, 
\)
where $k$ represents the three-dimensional momentum, and 
$n_i R^{-1}$ the integral internal momenta.\footnote{This spectrum
applies to a square torus with all radii equal to $R$. In
general there may exist other shape parameters that would change
the precise structure of the KK towers~\cite{Dienes}. 
However since we are only interested in generic effects associated
with the radion and gravitons, these effects only give subleading
corrections. Neither we consider more complicated models in which the KK 
spectrum does not depend on the size of the extra dimensions and very 
heavy KK modes can emerge from large extra dimensions~\cite{chm}.} 
Collectively, the square sum of these internal momenta
looks like a mass squared, denoted by $\mk^2$.

The coupling of each KK mode to matter is small, as it is proportional
to  $1/M_{\rm p}$. Moreover, the gravi-photon has no coupling to the
matter fields at the tree-level, and hence can be ignored in the
subsequent discussion. 
%The radion determines the volume of extra
%dimensions through its vacuum expectation value (VEV). 
The
coupling of a radion to fermions is proportional to the fermion mass
$m_f$; its coupling to massless gauge bosons vanishes at the 
tree-level.

The coupling, and therefore the emission probability, of a
particular KK mode is  very small 
due to the large size of extra dimensions as compared to the
electroweak scale. However, due to the small energy level spacing 
between the KK modes, the
multiplicity of kinematically allowed KK modes grows rapidly with
energy. The resulting probability of emission of KK modes scales like  
$(ER)^{d+2}E^2M^{-2}_{\rm p}\sim (E/M_*)^{d+2}$. This has an
important impact on particle physics through the production of on-shell  
and exchange of off-shell KK modes in various processes. 
Collider experiment constraints on the fundamental scale $M_*$ are 
comparable to 1 TeV scale~\cite{exp1}. Much more stringent bounds can
be derived from cosmology and astrophysics, as we review in section
\ref{sc:PreviousBounds}.

All cosmological bounds on the scale of extra dimensions derived in the 
literature rely on thermal production of the KK modes from the
scattering of SM particles. It is usually assumed that the temperature
is at its lowest possible value consistent with Big Bang
Nucleosynthesis (BBN)~\cite{olive}, i.e.\ $\ga 0.7$
MeV~\cite{bbnlow}. However, it is extremely problematic to start the
history of the Universe from a thermal state with such a low
temperature without having a pre-thermal history. Standard
cosmological problems such as flatness, isotropy, etc., prompt the
introduction of appropriate  inflationary models in the context of
extra dimensions ~\cite{dvali,kl,asymmetric,bulk}. Inflation in models
with large extra dimensions can be quite different from well-studied
inflationary scenarios in four dimensional theories. The inflaton has
to be a bulk field, since the possibility of a brane inflaton is not
really viable~\cite{kl}: for $M_{*} \sim {\cal O}({\rm TeV})$ this
requires very unnatural initial conditions, as well as an inflaton
which is too light to generate density perturbations of the correct
amplitude.   
The size of the extra dimensions needs to be stabilized before the
time of  BBN, otherwise the observationally tightly constrained
abundances of light elements ~\cite{bbn} will be distorted. We assume
that extra dimensions were somehow stabilized before, or during
inflation~\cite{mp}, so that for the last e-foldings of inflation,
relevant for the observable density perturbations, only three ordinary
dimensions are still expanding~\cite{bulk}. The requirement of
reheating before BBN imposes bounds on the width and the mass of the
inflaton field, with the latter being on the order or larger than the
electroweak scale $v_{EW}$. In addition, baryogenesis in models with
large extra dimensions~\cite{bd} also requires the existence 
of fields masses larger than $v_{EW}$~\cite{aemp}.

Therefore, in all realistic scenarios studied to date, the Universe had
to pass through a stage of entropy production in which the characteristic 
energy/momentum 
transfers were at least of the order of the electroweak scale. 
For simplicity we refer to the field $\phi$ which is responsible for this 
stage as the ``inflaton'' in
this work. Since gravity is the universal force which couples to all
energy sources, the decay of the inflaton is necessarily accompanied
by non-thermal emission of KK gravitons with masses comparable to
$m_\phi$. The resulting KK excitations may survive for several years,
and decay when all light elements have already been formed. Their
decay products may distort the abundance of light elements predicted
from BBN. Consequently, as we will quantify in this paper, these
non-thermal processes can strengthen the cosmological bounds on 
large extra dimensions, especially if $d \ga 3$.

This paper is organized as follows. Section \ref{sc:PreviousBounds} 
reviews previously obtained bounds on the scale of the large extra
dimensions using astrophysical and cosmological probes. In section
\ref{sc:InflatonDecay} we first argue that non-thermal production of KK
modes is inevitable in models with large extra dimensions because of
the need of bulk inflation. After that, we identify non-thermal
processes of KK mode production, like the decay of a heavy inflaton,
and investigate their potential to constrain $M_*$ in various extra
dimensions. Our main bounds can be found in Table \ref{tab:bounds}. 
Section \ref{sc:Concl} summarizes our conclusions. The calculations of
the emission of  KK radions and gravitons in the decay of the inflaton
are summarized in Appendix \ref{sc:KKEmission}.

\section{Cosmological bounds on the scale of extra dimensions
via thermal production of KK modes}            
\label{sc:PreviousBounds}

This section gives a short overview of the various bounds
on $M_*$, as a function of the number of extra dimensions $d$, that
have been obtained from various cosmological and astrophysical
considerations. We also briefly indicate what kind of physics has been
employed to obtain these bounds.

After reheating $T< T_{\rm R}$, the Universe should be described by a
thermal bath, which is radiation-dominated at least during BBN, 
$T_{{\rm R}} \ga 0.7$ MeV~\cite{bbnlow}; otherwise the well-measured
abundances of the light elements become distorted. However, since the
massive KK modes behave as non-relativistic matter, their existence
may be problematic. First of all, above the so-called {\em normalcy} 
temperature $T_{\rm N}^{d+1} \sim M_*^{d+2}/M_{\rm p}$ the KK modes dominate,
hence one requires that $T_{{\rm R}} < T_{\rm N}$~\cite{nima0}. In
addition, KK modes redshift more slowly than radiation, and therefore
will become dominant eventually, provided that their lifetime is
sufficiently long. This results in a tight bound on $M_*$ from the
over-closure limit. For example, for $d=2$ and $T_{{\rm R}} \gs 3$
MeV, one finds  $M_* \gs 40$ TeV~\cite{hs}. This bound on $M_*$ can be
improved by considering the diffuse photon background, produced in the
decay of a KK graviton into a pair of photons in the MeV range~\cite{hs}. 
This diffuse gamma background has been measured by the
EGRET~\cite{egret} and COMPTEL~\cite{comptel} experiments, and from 
this the bound $M_* \ga 110$ TeV for $d=2$ follows~\cite{hs}.

Even more stringent bounds can be obtained from KK modes, 
which are thermally produced during reheating. 
In this epoch the Universe is matter-dominated and an instantaneous
thermal bath exists~\cite{kt}, with maximum temperature 
$T_{max}$~\cite{ds,ad2},  which carries a small, but growing, fraction of the 
total energy density. 
Due to the continuous injection of entropy from inflaton decay, $T$
decreases more slowly than if radiation were dominant. As a consequence, 
the number density of produced particles is redshifted more rapidly as
a function of temperature, i.e.\ $\propto T^8$, instead of  
$\propto T^3$ for radiation~\cite{kt}. The KK modes with masses  
$T_{{\rm R}} < \mk < T_{max}$ are produced with only a power-law
suppression during reheating, rather than an exponential suppression in a
radiation-dominated Universe. This has been used in Ref.~\cite{h} to
tighten the bound on $M_*$. In fact, with the lowest BBN allowed
reheat temperature, $T_{\rm R} \approx 0.7$ MeV and  
$T_{max} = 1$ GeV, a bound of $M_{*} \gs 167~(21.7)$ TeV has been 
obtained for $d=2~(3)$.

The KK modes production in hot and dense astrophysical objects also leads 
to very tight bounds on $M_*$~\cite{hr1,hr2}. For $2$ extra
dimensions one finds in this way $M_* \gs 3900$ TeV~\cite{hr2}.

\begin{table}
\begin{center}
\renewcommand{\arraystretch}{1.5}
\begin{tabular}{|c||c|c|c|c|c|}
\hline 
Bounds from &  $d=2$ & $d=3$ &$d=4$&$d=5$& $d=6$
\\ \hline \hline
cosmology & 167 TeV & 21.7 TeV & 4.75 TeV & 1.55 TeV 
& $< 1$  TeV${}^\dag$
\\
astrophysics & 3930  TeV & 146  TeV & 16.1 TeV 
& 3.4 TeV & 1 TeV
\\ \hline 
\end{tabular}
\end{center}
\caption{\label{tab:currentbounds}
The tightest current cosmological~\cite{h} and astrophysical~\cite{hr2} 
bounds on the Planck scale in $4+d$ dimensions 
$M_*$, for $2\leq d \leq 6$. Note that with our defintion of the fundamental 
scale, Eq.\ \eqref{volume}, $M_*$ coincides with $M$ in Ref.~\cite{h},
while related to $M$ in Ref.~\cite{hr2} through the relation 
$M_* = (4\pi)^{1/d+2} M$. ${}^\dag$ A cosmological bound for $d=6$
has not been given in~\cite{h}; but it is certainly less than 1 TeV.}
\end{table}

Both the cosmological and astrophysical bounds rely on the thermal 
production of KK modes at temperatures in the range of $1$ to $100$
MeV. The probability of a KK mode production in a single fusion
process scales as $(T/M_*)^{d+2}$; hence for larger $d$ the bounds
derived from such processes do not give very stringent constraints. 
The strongest cosmological and astrophysical bounds, which are
currently available,  have been listed in Table \ref{tab:currentbounds}.

\section{Non-thermal production of KK modes in inflaton decay}
\label{sc:InflatonDecay}

As recalled in the introduction, the existence of a bulk inflaton is a
fairly generic feature of viable inflationary scenarios in large extra
dimensions. This inflaton has to produce the SM fields, which
live on the brane, in order to reheat the Universe. This can either be
a direct process, or the inflaton first decays to other bulk or brane
fields, which in turn decay to the SM particles. In either case, the
reheat temperature $T_{\rm R}$ must be sufficiently low to avoid 
thermal overproduction of KK modes, as mentioned in section
\ref{sc:PreviousBounds}.  A higher $T_{\rm R}$ would require a 
larger $M_*$, or late entropy release in order to dilute the KK modes in 
excess. Note that the energy density of a brane field is in general 
too small to accomplish this. In conclusion, the situation is that 
the decay of a bulk field will typically be responsible for the final 
stage of reheating, which yields $T_{\rm R} \simeq 1$ MeV.

The decay rate of the inflaton depends on its coupling to the SM
fields. The decay to the Higgs, to fermions, and to gauge fields
corresponds to coupling through a dimension $D = 3, 4,$ and $5$
operator, respectively. After taking into account the volume suppression
factor,  the resulting decay widths scale like 
\be \label{decayD}
{\Gamma}_{D} \sim \frac {M_*^3}{M_P^2} 
\Bigl( \frac {m_{\phi}}{ M_{*}} \Bigr)^{2D-7}. 
\ee
(The minimum $D=3$ corresponds to a $\phi {\rm H}^\dagger {\rm H}$ 
operator, in which case Eq.\ (\ref{decayD}) is applicable only if
$m_\phi > 2 m_H$.) Recall that for successful BBN, the total decay
width $\Gamma_{tot}$ of the inflaton has to satisfy 
${\Gamma_{tot}} \sim T^{2}_{\rm R}/M_{\rm P}$, with the reheat
temperature $T_{\rm R} \ga 0.7$ MeV. With $M_{*}$ not hierarchically
larger than 1 TeV, this implies that $m_\phi$ must be of the order of
1 TeV, as well.  Note that this is, in fact, a generic conclusion for
the decay of any bulk field into fields localized on the brane.

\begin{figure}[t]
\begin{center}
\epsfig{file=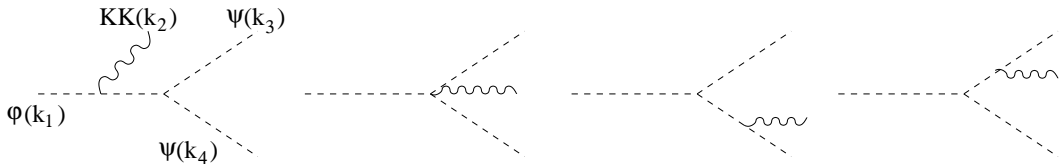, width=14cm, angle=0}
\end{center}
\vskip .1in
\caption{\label{fig:3body} 
The associated emission of KK modes in the decay of the inflaton into
a pair of the scalars $\psi$ from the incoming leg,  the vertex, and
the outgoing legs. }
\end{figure}

The universal character of the gravitational interaction ensures that
any decay of the inflaton $\phi \rightarrow \psi_i \psi_j$ to the SM
fields $\psi_i$ is accompanied by KK mode emitting processes 
$\phi \rightarrow \psi_i \psi_j+KK$ with $\mk\la m_{\phi}$.\footnote{In 
general, particles with a mass $\ls m_\phi$ are 
produced in inflaton decay in higher orders of perturbation theory whenever 
they can be produced via thermalized inflaton decay 
products~\cite{ad1}.} These KK
modes may carry a considerable fraction of the energy of the decaying
inflaton. Let us concentrate on the inflaton decay into the Higgs scalars. As
we shall see below, the coupling constant of the inflaton to matter
does not enter in the expression for the total energy emitted to KK
modes, and hence our analysis is rather insensitive to the nature of
the SM final states, and their coupling to the inflaton. The diagrams
describing the emission of KK radions and gravitons in the decay of
the  inflaton to two scalars are depicted in Fig.\ \ref{fig:3body}.  
(KK modes with ${\cal O}(100~\rm GeV)$ masses are also produced during
reheating, provided that  $T_{max}$ is sufficiently large. However,
compared to those produced in inflaton decay, their abundances are
negligible, as they are suppressed by large powers of  
$T_{\rm R}/\mk$.) 

For the KK modes decaying after recombination, i.e. 
${\tau}_{KK} > 10^{12}$ sec, the gamma ray background provides the
strongest bound on the total energy $\sum n_{KK}\cdot \mk/s$
released in KK modes~\cite{sarkar,kr}.  (Here $\sum$
denotes the sum over all KK modes which are kinematically allowed.) 
For shorter lifetimes, ${\tau}_{KK} < 10^{12}$ sec, the constraints
from dissociation of primordial light elements by KK mode decay
products are stronger~\cite{sarkar,cefo}. The most stringent BBN bound
arises for $10^{8}$ sec $\la {\tau}_{KK} \la 10^{12}$ sec, and
reads~\cite{cefo} 
\be \label{bbnbound}
\frac {\sum n_{KK}\cdot \mk}{ s\cdot 1\,{\rm GeV}} \simeq 2\cdot 
10^{-12}.
\ee

\begin{figure}[t]
\begin{center}
\raisebox{0ex}{\scalebox{0.7}{\mbox{\input{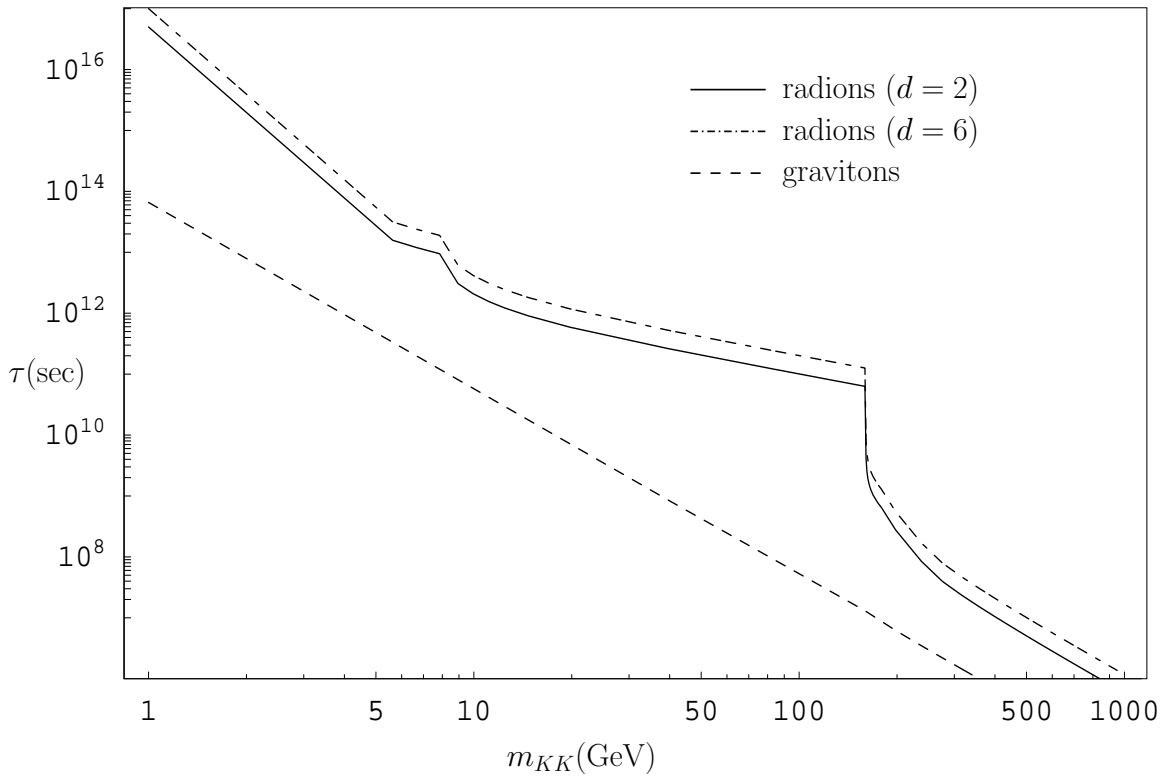}}}}
\end{center}
\caption{\label{fig:lifetimes} 
The lifetime of the KK gravitons as a function of their mass is
indicated by the dashed line. The lifetime dependence of the KK
radions on the number of extra dimensions is shown by the solid line
for $d=2$ and the dashed-dotted line for $d=6$. In addition, the
radion lifetime is very sensitive to $m_{KK}$ near the $WW$ threshold.
} 
\end{figure}

The energy emitted into KK modes with even shorter lifetimes, 
$10^{6}$ sec $\la {\tau}_{KK} \la 10^{8}$ sec, is bounded by the 
BBN predictions at the level of $2\cdot 10^{-9}$ \cite{cefo}. 
Such lifetimes correspond to masses of KK modes comparable to the
electroweak scale. Indeed, KK gravitons with a mass larger than the 
electroweak scale decay to all SM particles with a decay 
lifetime~\cite{hlz}
\be \label{gravitondecay}
{\tau}_{\rm gr} \simeq  5 \times 10^{4}~{\rm sec} 
\left(\frac{1~ {\rm TeV}}{\mk}\right)^3,
\ee
while the lifetime of the radions, capable to decay into  
$ {\rm H} {\rm H},~W^+W^-,~Z Z$, is 
given by~\cite{hlz}
\be \label{radiondecay}
{\tau}_{\rm rad} \simeq 2(d+2)\times 10^{5}~{\rm sec} 
\left(\frac{1~{\rm TeV}}{\mk} \right)^3 
%~~{\rm for}~\mk\ga 200~ {\rm GeV}. 
\ee        
The lifetime of massive radions is substantially longer than that of
massive gravitons, because the latter has a lesser number of efficient
decay channels. The $\tau_i$ dependence on the KK mass $\mk$ is
plotted in Fig. 2.  Assuming that the decaying inflaton is heavier than
all massive KK excitations, 
\be
m_{max}^{\rm gr} \simeq 80 {\rm GeV}; 
\qquad 
\begin{array}{l c c c c c c l }
m_{max}^{\rm rad}  & \simeq &  200, & 210, & 225, & 240, & 250 &
{\rm GeV},  
\\[.3ex] 
\text{for}\, ~ d & = & 2, & 3, & 4, & 5, & 6, & 
\end{array}
\label{mmax}
\ee
are the maximum masses of the KK gravitons and radions, respectively,
which give the lifetime of $10^{8}$ sec and allow to use the tight bound 
(\ref{bbnbound}). 
%lifetimes of these KK modes to $10^{8}$ sec.

\begin{figure}[t]
\begin{center}
\epsfig{file=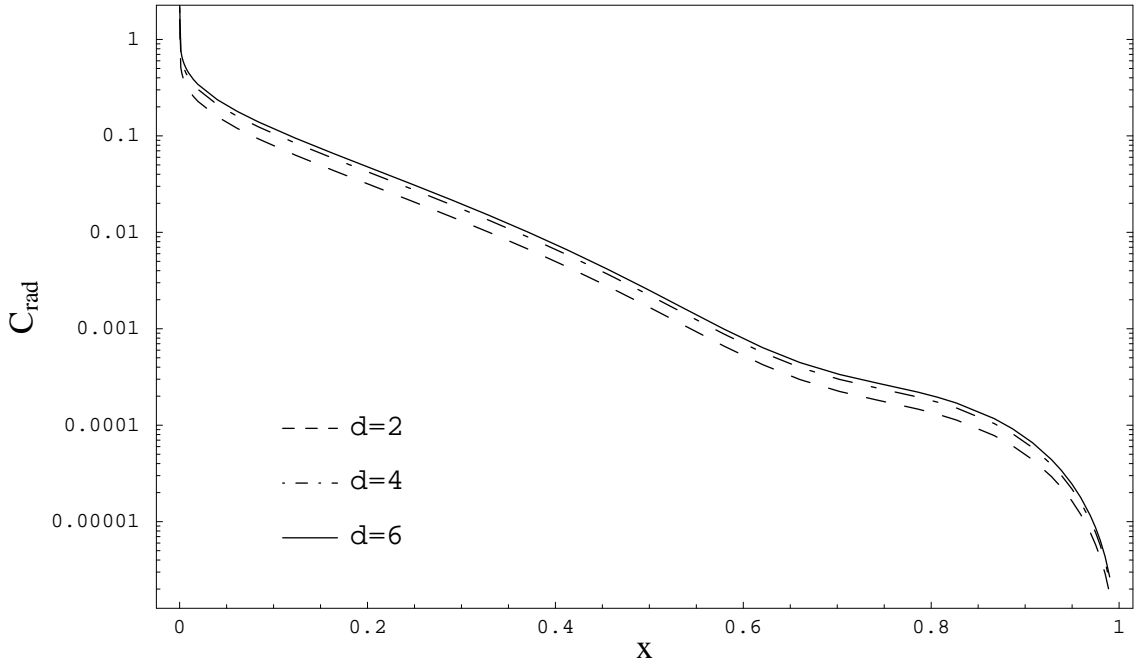, width=15cm, angle=0}
\end{center}
\caption{\label{fig:LogC} 
The dependence of the factor $C_{{\rm rad}}(x)$, defined
in \eqref{decaykk}, on the ratio $x = m_{KK}^{\rm rad}/m_{\phi}$ of
the radion over the inflaton mass is plotted for the number of extra
dimensions $d= 3,4,$ and $6$. }
\end{figure}

We now compute the energy density carried by radions and gravitons
with $\mk \leq m_{max}$ based on decay width calculations presented in
Appendix \ref{sc:KKEmission}. The ratio of the number density of
emitted single KK mode over the number density of the inflaton
$n_\phi$, can be expressed in terms of a function $C_i(x)$ as 
\be \label{decaykk}
\frac{n^i_{KK}}{n_{\phi}} = C_i (x)  
\frac{m_{\phi}^2}{M_{\rm P}^2},
\qquad 
x = \frac{\mk^i}{m_\phi}, 
\qquad 
i = {\rm gr}, ~{\rm rad}.
\ee
The functions $C_i(x)$ depend on the dimension $d$, and are different
for radions and gravitons, and have been computed in Appendix
\ref{sc:KKEmission}. In particular, their asymptotic expressions have 
been given in \eqref{Crad} and \eqref{Cgr} of this Appendix. 
In Fig.\ \ref{fig:LogC} we have plotted $C_{{\rm rad}}(x)$ for 
$d = 2, 4,$ and $6$. 
\begin{figure}[t]
\begin{center}
\epsfig{file=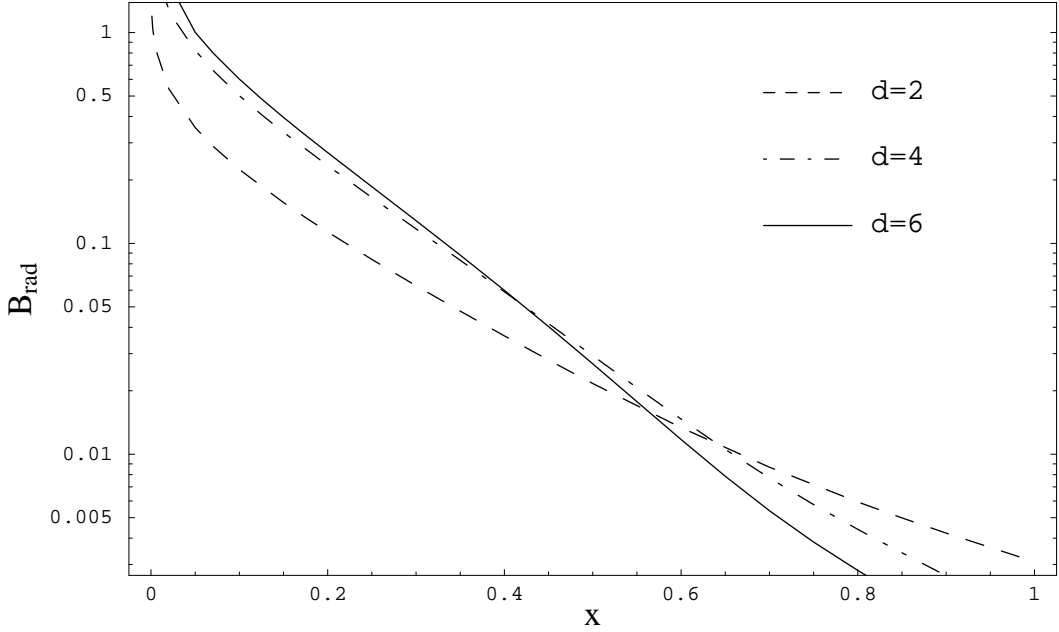, width=14cm, angle=0}
\end{center}
\caption{\label{fig:LogB} 
The dependence of $B_{{\rm rad}}(x)$, given in 
\eqref{final}, on the ratio $x = m_{max}^{\rm rad} / m_{\phi}$ is
presented for the number of extra dimensions $d = 2,4,$ and $6$.} 
\end{figure}

By combining the expressions for the number densities (\ref{decaykk}) 
with the entropy released by inflaton decay 
\be \label{entropy}
\frac{n_{\phi}}{s} \simeq \frac{3 T}{m_{\phi}},
\ee
we arrive at the expression for the energy density of KK modes
weighted with entropy factor
\be \label{final}
\frac{1}{s} \,  \sum \mk^i n^i_{KK} \simeq 
3 B_i(x) 
\frac{ T m_\phi}{M_*} 
\Bigl(  \fr{m^i_{max}} {M_*} \Bigr)^{d+1},
\qquad 
x = \frac {m^i_{max}}{m_\phi}.  
\ee
Here the functions $B_i(x)$ have been obtained by summing 
$\sum \mk^i C(\mk^i/m_\phi )$ over all KK modes with a mass smaller
than the maximal KK mass $m_{max}^i$. Since the spacing of modes is
$1/R$, we may replace this sum by an integral over a sphere of radius
$Rm_{max}^i$, which, after some change of variables, becomes  
\be
B_i (x)  = 
\frac { S_d}{x^{d+1}} 
\int_0^{x} d y\, 
C_i(y)\, y^{d}, 
\qquad 
x = \frac {m_{max}^i}{m_\phi},
\ee
where $S_d=2\pi^{d/2}/\Gamma(d/2)$ is the surface area of the 
$d$-dimensional unit sphere.  
For the radion and $d = 2,~4,$ and $6$, we have plotted the function
$B_{{\rm rad}}(x)$, see Fig. \ref{fig:LogB}. In the limit 
$x = m_{max}^i/m_\phi \ll 1$, $B_i(x)$ and $C_i(x)$ are related by
$B_i(x) \sim C_i(x) S_d/(d+1)$. The expressions in \eqref{final} remain
approximately constant in the subsequent thermal evolution, since the
entropy and KK number densities redshift in the same manner.

\begin{table}
\begin{center}
\renewcommand{\arraystretch}{1.5}
\begin{tabular}{|c||c|c|c|c|c|}
\hline 
$m_\phi$ & $d=2$ & $d=3$ &$d=4$&$d=5$& $d=6$
\\ \hline \hline
1 TeV & 35 TeV  & 13 TeV & 7.1 TeV & 4.5 TeV & 2.8 TeV
\\ 
2 TeV & 47 TeV  & 17 TeV & 9.1  TeV & 5.7 TeV & 3.4 TeV
\\
$M_*$ & 220 TeV  & 42 TeV &  15 TeV &
7.9 TeV &  4.0 TeV
\\ \hline
\end{tabular}
\caption{\label{tab:bounds}
The cosmological bounds on the Planck scale in $4+d$ dimensions 
$M_*$, for $2\leq d \leq 6$ and inflaton masses $m_\phi = 1,~2$ TeV and
$M_*$, are obtained by requiring that late decay of KK excitations do
not distort the BBN predictions. These bounds can be compared with 
the previous bounds collected in Table \ref{tab:currentbounds}.
}
\end{center}
\end{table}

From now on we assume, that $T \simeq  T_{{\rm R}} \approx 0.7$ MeV, 
since higher reheat temperatures only lead to stronger bounds. 
In the asymptotic regime of $x = m_{max}/m_\phi \ll 1$ the average 
non-relativistic energy density carried by the sum of KK radions and
gravitons can be written as 
\be
\frac{\sum \mk n_{KK} }{s} 
\simeq 
\frac{2\, S_d}{(d+1)\pi}
\frac{T_{\rm R}m_\phi}{M_*}
\left[
\frac {d}{d+2} 
\Bigl( \fr{m_{max}^{\rm rad}}{M_*} \Bigr)^{d+1} 
+ 
2 \Bigl( \fr{m_{max}^{\rm gr}}{M_*} \Bigr)^{d+1} 
\right]
\ln \Bigl( \fr {m_\phi}{m_{max}} \Bigr), 
\label{naive}
\ee
where we have used the approximation 
$\ln m_{max}^{\rm rad}\simeq \ln m_{max}^{\rm gr} $.  
Imposing the constraint (\ref{bbnbound}) on (\ref{naive}), we arrive at a set 
of bounds on $M_*$, which are summarized in the Table \ref{tab:bounds}. 
The strength of these limits is mainly due to $200$ GeV radions. The 
contribution of the gravitons is small, and does not exceed 15\% of the total 
$\sum\mk n_{KK}$, because of a factor of $3$ difference in
the maximum values of $\mk$ for radions and gravitons, given in 
(\ref{mmax}). 
Below \eqref{bbnbound} we noted that for lifetimes of $10^{6}$ sec 
$\la {\tau}_{KK} \la 10^{8}$ sec of KK modes, the BBN bound is
weakened to $2\cdot 10^{-9}$ \cite{cefo}. However, using these numbers
instead, does not affect the limit on $M_*$ quoted in Table
\ref{tab:bounds} much for $m_\phi =1$ TeV. For heavier inflatons the
limits would be strengthened. For example, for $d=6$ and $m_\phi = 2$
TeV,  the bound on $M_*$ is at the 4 TeV level. 

It is important to note that in the approximation $m_{max}\ll m_\phi$
the first diagram in Fig.\ \ref{fig:3body} dominates and contributes
to Eq.\ (\ref{naive}). Therefore, in this approximation, the reheating
to other particles would produce the same answer. We checked that this
is indeed the case for the reheating into a pair of fermions. Thus,
our bounds are practically insensitive to the nature of the
inflaton-matter coupling. 

Associated inflaton decay into KK modes calculated in this 
section is not the only source of non-thermal production of radions
and gravitons. KK modes will also be produced during the
thermalization process of the inflaton decay products. In this process
energetic inflaton decay products scatter among themselves and with
the particles in the thermal bath, which have energies of the order of
the reheat temperature $T_{\rm R}$~\cite{ad2}. The scattering of energetic
particles with energies $E\sim m_\phi/2$ would lead to a production of
KK modes with masses up to  $\mk \ls m_{\phi}$. An estimate of
$s^{-1}\sum \mk n_{KK}$ produced this way shows that it is subdominant
to (\ref{final}). 
The main reason for this is, that the scattering of the inflaton decay 
products is suppressed by the square of rather small number 
density of energetic particles. This leads to additional suppression 
by a square of the small parameter $ T_{\rm R}/m_\phi$.
The scattering of the hard particles on the particles 
from the thermal bath may produce KK modes with masses 
$\mk \ls (Tm_\phi)^{1/2}$. This domain of KK masses is rather small
to be important for the production of the KK modes in $d>3$ models. 
Estimates of the energy densities of the KK modes producing
the diffuse photon background for $d=2,~3$ models 
show that the resulting level of sensitivity to $M_*$ is lower than 
previously found in Refs. \cite{hs,h}. It is interesting to 
note in passing that the largest production of KK modes during thermalization 
is expected from the thermalization of neutrinos because they 
interact via the weak force with the thermal bath and thus 
exist much longer in this environment than the rest of the particles.

\section{Conclusions}
\label{sc:Concl}

We have presented a derivation of cosmological bounds on the 
fundamental scale of quantum gravity $M_*$ in large extra dimension
models. These bounds are obtained by utilizing the non-thermal 
production of massive KK gravitons and radions from
the decay of a heavy inflaton field. Such a heavy field seems to be a
generic feature of models with large extra dimensions: 
Inflationary resolutions of standard cosmological problems together
with successful reheating require, that the inflaton mass is larger than
the electroweak scale.  Moreover, it seems to be impossible to
generate the baryon asymmetry of the Universe without invoking some 
physics beyond the electroweak scale. Our limits on $M_*$ arise 
because the subsequent decay of those KK modes after the BBN epoch may
distort the abundances of light elements, which are tightly
constrained by observations.

The production of KK modes during inflaton decay cannot be computed 
in a fully model-independent way. However, the (potentially) most
uncertain parameter, the inflaton coupling to the matter, dropped out of
the calculation of the total energy density of KK modes. Also the
reheating into fermions or gauge bosons, instead of only scalars as we
considered, will not drastically change the conclusions from the
calculation. This possibly leads to an order unity change in numerical 
coefficients, but the dependence on the main parameters $m_{max}$,
$m_{\phi}$ and $M_{*}$ remains the same as in (\ref{naive}).

Our limits on the fundamental scale $M_*$ have been summarized in
Table \ref{tab:bounds}: For {\em any} number of dimensions the resulting
limits are stronger than 2 TeV. Hence, they are clearly superior to 
existing high-energy physics bounds, and very competitive with any 
possible level of sensitivity to $M_*$ of future collider experiments
of this decade. For a TeV scale inflaton the cosmological bounds found in the
literature  ~\cite{hs,h} are stronger than ours for $2$ or $3$
extra dimensions. However, our limits can become more 
stringent for a heavier inflaton. Indeed, the saturated bounds in Table 2, 
obtained for $m_\phi = M_*$, are certainly tighter than the existing ones. 
For $m_\phi \geq 1$ TeV, our limits are better than the current
strongest one for $d=4$~\cite{h}, and are clearly an improvement for
$d =5$ and $6$. The most stringent astrophysical bounds which currently
exist~\cite{hr2} are much stronger than our bounds for $d=2,~3$, and are 
better for $d=4$. However, our bounds are again superior for $d=5,~6$.

The reason that our bounds are particularly powerful for
large $d$ is the following. The existing cosmological bounds rely on 
thermal production of KK modes, which have much smaller
$\mk$. Therefore, even though their abundances are enhanced  
by a large time of emission, for sufficiently large $d$ the phase
space factor takes over, and results in a larger abundance for
non-thermally produced KK modes from inflaton decay.  

\section*{Acknowledgements}

R.A.\ and S.G.N.\ wish to thank kind hospitality by the Department of
Physics and Astronomy at the University of Victoria, and the Theory
group of TRIUMF. M.P.\ would like to thank I.\ Mocioiu for useful
discussions at the  initial stage of this project. 
\\
This work has been supported by the NSERC of Canada. In addition, 
S.G.N.\ acknowledges the National Fellowship support of CITA, and 
M.P.\ the support of PPARC of the UK.

\appendix
\section{Emission of KK modes}
\label{sc:KKEmission}

In section \ref{sc:InflatonDecay} we discuss how the inflaton decay
to KK gravitons and radions can be used to put bounds on the
fundamental $4+d$ dimensional Planck scale $M_*$. This appendix is
devoted to the calculation of an estimate of the ratio of the number density of
produced KK modes $n^i_{KK}$ to the number density of the inflaton
$n_{\phi}$ on which these bounds crucially rely. This ratio is equal
to the three body decay width containing a single KK mode
${\Gamma_{3\, i}}$ over the total decay width ${\Gamma_{tot}}$ of the
inflation:  
\be 
C_{i}(x)
\frac{m_{\phi}^2}{M_{\rm P}^2}
\equiv 
\frac {n_{KK}^i}{n_\phi} = \frac{\Gamma_{3\, i}}{\Gamma_{tot}} 
\simeq 
\frac {\Gamma_{3\, i}}{\Gamma_2}, 
\quad 
x = \frac{\mk^i}{m_\phi},  
\quad 
i = {\rm rad}, ~{\rm gr}.
\ee
Here we have assumed that the two body decay width $\Gamma_2$ of the
inflaton to SM particles dominates its total width $\Gamma_{tot}$. 
(As can be seen from \eqref{decayD} generically a two body decay (into
scalar) gives the dominant contribution to the total decay width.)
As discussed in section \ref{sc:InflatonDecay}, we further estimate
all relevant SM processes by the decay of the inflaton to scalars
$\psi$. For this we assume the simple model 
\begin{equation}
{\cal L} = 
\fr{1}{2} \partial_{\mu} \phi \partial^{\mu} \phi -\frac{ m_{\phi}^2}{2} 
\phi^2 + 
\fr{1}{2}\partial_{\mu} \psi \partial^{\mu} \psi -\frac{ m_{\psi}^2}{2} 
\psi^2+ 
\frac{\lambda }{2}\phi 
\psi^2.
\end{equation}
Moreover, in the physical situations of interest in the main text 
 $m_\phi \gg m_\psi$, hence we consider the massless limit for
$\psi$. The width of the two body decay $\phi \to \psi\psi$ is then
given by 
\begin{equation}
\Gamma_2 = \frac{\lambda^2}{32 \pi m_{\phi}}.
\end{equation}
The relevant diagrams for the three body decays involving a KK mode are
given in Figure \ref{fig:3body}, and we define $\Gamma_{3,i}$ as
% result in the usual phase space integral 
%
\begin{equation}
\Gamma_{3 , i} = 
\int \frac{d|M_{i}|^2}{128 \pi^3 m_{\phi}} dE_2 \; dE_3, 
\qquad 
i = {\rm rad}, ~{\rm gr}. 
\end{equation}
For the estimates of the amplitudes, we need to
make a distinction between KK radions and gravitons.

The amplitude for the radion producing reaction 
$\phi (k_1) \to {{\rm rad}} (k_2) + \psi (k_3) + \psi(k_4)$ reads 
(in the same ordering as the diagrams given in Fig.\ \ref{fig:3body})
\begin{equation}
iM_{{\rm rad}} = -i \kappa \omega \lambda \left( \frac{k_1 \cdot k_2 + 
m_{\phi}^2}{2 k_1 \cdot k_2 - \mk^2} -2 + \frac{k_2 \cdot k_3}{2 k_2 \cdot 
k_3 + \mk^2}+ 
\frac{k_2 \cdot k_4}{2k_2 \cdot k_4 + \mk^2} \right),
\end{equation}
where we have used $\kappa^2M_{\rm p}^2=16\pi$ and the normalization 
factor $\omega^2 = \smash{\frac{2}{3(d+2)}}$ of the radion
~\cite{hlz}. By computing the relevant phase space integral and summing 
over $d$ identical radions, we can
finally obtain the full expression for $C_{\rm rad}(x)$. Since its
form is rather complicated, we have plotted this function for 
$d=3,4,$ and $6$ in Fig.\ \ref{fig:LogC}. Its asymptotic expression
reads 
\be 
C_{{\rm rad}}(x) \simeq \frac{2d}{3\pi(d+2)} \ln \Bigl( \frac 1x \Bigr),  
\qquad 
\qquad 
\text{for} 
\qquad 
x = \frac{\mk^{\rm rad}}{m_\phi} \ll 1.
\label{Crad}
\ee

The matrix element for graviton production 
$\phi(k_1) \to {\rm gr} (k_2) + \psi(k_3) + \psi(k_4)$ can be obtained
in a similar fashion using the diagrams of Fig.\
\ref{fig:3body}. However the amplitude is now a tensor
quantity. Furthermore, since $m_{max}^{{\rm gr}}$ is significantly
smaller than $m_{max}^{{\rm rad}}$ (see \eqref{mmax}) we may neglect
the masses of the KK gravitons. The amplitude for the KK gravitons
then becomes   
\begin{align}
iM^{\rm gr}_{\mu \nu} & =  i \frac{\kappa \lambda}{2} \left(
- \frac{m_{\phi}^2 \eta_{\mu \nu} 
+ C_{\mu \nu \rho \sigma} k_1^{\rho} (k_1^{\sigma} - k_2^{\sigma})}
{2 k_1 \cdot k_2}
+
\frac{C_{\mu \nu \rho \sigma}(k_3^{\rho} +k_2^{\rho})k_3^{\sigma}}
{2 k_2 \cdot k_3}
\right.
\\
& 
\left.
 + 
\frac{C_{\mu \nu \rho \sigma}
(k_4^{\rho}+k_2^{\rho})k_4^{\sigma}}{2 k_2 \cdot k_4}+\eta_{\mu \nu} \right),
\quad \text{with}\quad 
C_{\mu \nu \rho \sigma} = \eta_{\mu \rho} \eta_{\nu \sigma} + 
\eta_{\mu \sigma} \eta_{\nu \rho} - \eta_{\mu \nu} \eta_{\rho \sigma}.
\nonumber
\end{align}
This gives for the KK gravitons in the massless approximation 
\be
C_{{\rm gr}}(x) = \frac{4}{3\pi} \ln \Bigl( \frac 1x \Bigr)
\qquad 
x = \frac{\mk^{\rm rad}}{m_\phi} \ll 1.
\label{Cgr}
\ee
Here we get a $\log(m_\phi/E) $ dependence, where $E$
 is the minimum energy carried by the graviton. To avoid an infrared
singularity, we assumed that the graviton has some minimum
energy, which is taken to be the mass $\mk^{\rm gr}$.

\end{document}